\documentclass[preprintnumbers,showpacs,twocolumn,secnumarabic,amssymb,amsmath,amsfonts,aps,prl]{revtex4}
\usepackage{amssymb,amsbsy,epsfig,color,graphicx}

\newcommand{\be}{\begin{equation}}
\newcommand{\ee}{\end{equation}}
\newcommand{\bea}{\begin{eqnarray}}
\newcommand{\eea}{\end{eqnarray}}

\begin{document}
\title{Modular Entanglement}
\author{Giulia Gualdi}
\author{Salvatore M. Giampaolo}
\author{Fabrizio Illuminati}
\affiliation{Dipartimento di Matematica e Informatica,
Universit\`a degli Studi di Salerno, Via Ponte don Melillo,
I-84084 Fisciano (SA), Italy; CNR-SPIN, and INFN Sezione di Napoli,
Gruppo collegato di Salerno, I-84084 Fisciano (SA), Italy}
\date{November 7, 2010}
\begin{abstract}
%We discuss possible ways and schemes to realize and distribute
%entanglement between distant and non-interacting quantum systems.
%We discuss the properties of renormalization, amplification, and scalability of long-distance entanglement
%in quantum many-body systems. From this investigation there emerges naturally the concept of modular %entanglement.
We introduce and discuss the concept of modular entanglement. This is the entanglement that is established between the end points of modular systems composed by sets of interacting moduli of arbitrarily fixed size. We show that end-to-end modular entanglement scales in the thermodynamic limit and rapidly saturates with the number of constituent moduli. We clarify the mechanisms underlying the onset of entanglement between distant and non-interacting quantum systems and its optimization for applications to quantum repeaters and entanglement distribution and sharing.
%In this paper we show that the class of models exhibiting  LDE is far wider than hitherto known.
%In facts, taking as constituting modulus an even site system capable of generating maximal end-to-end %entanglement, a composite system made up of identical interacting replicas of such a modulus exhibits extremal %quantum correlations which are non vanishing in the thermodynamic limit. This result provides a deep insight  %and paves the way to the possibility of long-distance entanglement optimization.
\end{abstract}
\maketitle
\begin{figure}[t]
{\includegraphics[width=6.3cm,height=2.4cm]{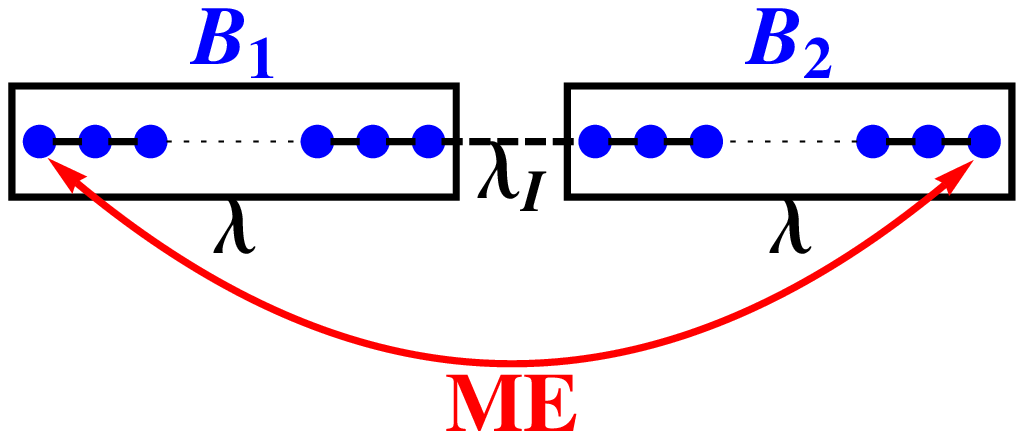}\vspace{0.2cm}
\includegraphics[width=6.3cm,height=2.4cm]{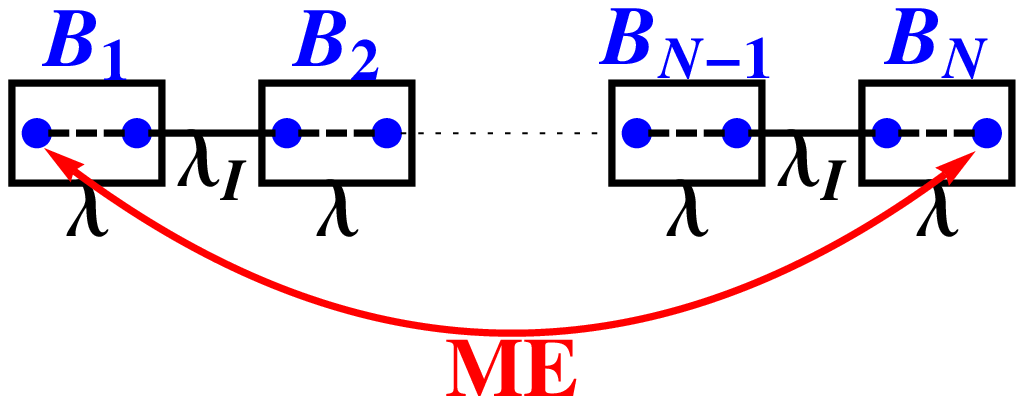}}
\caption{\label{Due-N-blocchi} (color online) Upper panel: A two-moduli system 
with end-to-end ME. The moduli interact via the local coupling $\lambda_I$. 
Inside each modulus, the bulk is uniformly interacting, whereas $\lambda$ denotes 
the weak end bond coupling. Lower panel: Generalization $N$ identically interacting moduli.}
\end{figure}
\begin{figure}[t]
{\includegraphics[width=7.5cm,height=3.8cm]{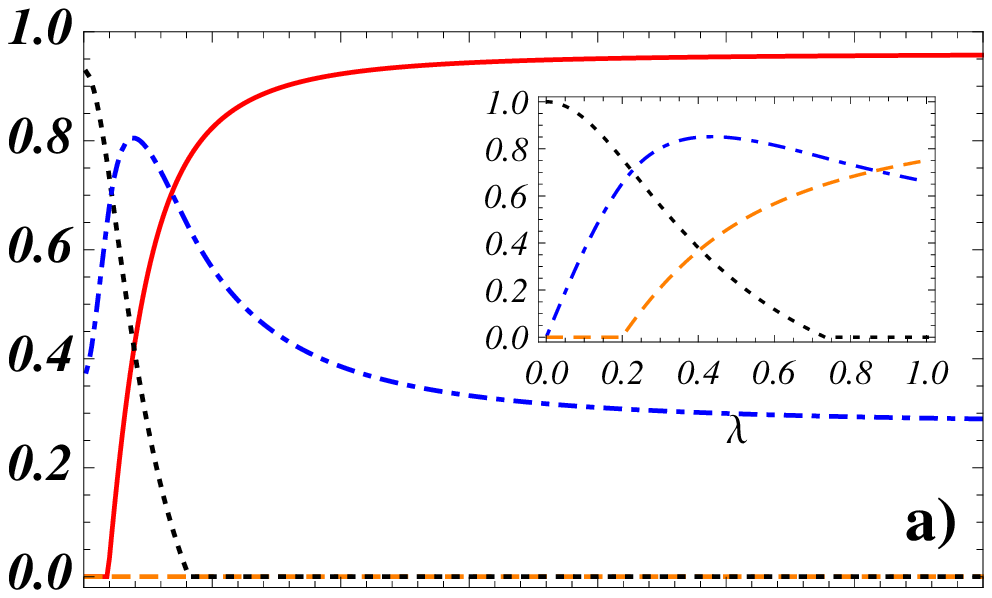}\vspace{-.35cm}
\includegraphics[width=7.5cm,height=3.8cm]{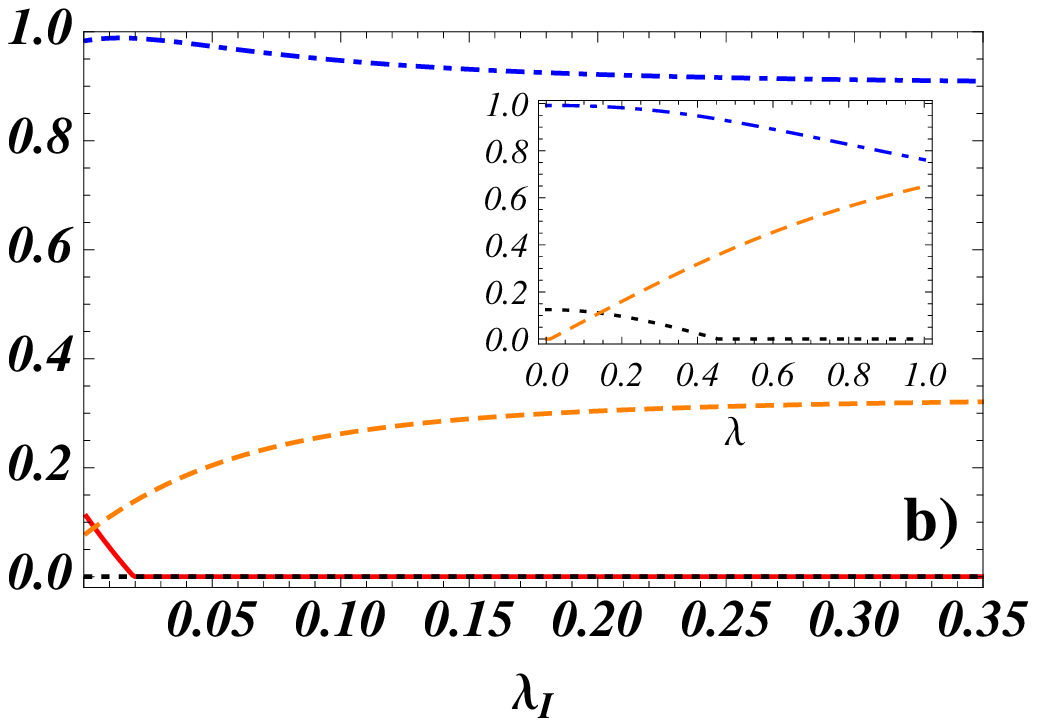}}
\caption{\label{2blocksevodd} (color online) Entanglement in a two-moduli spin system. Overall end-to-end ME $C_{1,2n}$ (solid red line), end-to-end entanglement inside a single modulus $C_{1,n} \equiv C_{n+1,2n}$ (dotted black line), end-pair concurrence $C_{1,2} \equiv C_{2n-1,2n}$ (dashed orange line), and multipartite entanglement $(\tau^{res}_{1})^{1/2}$ (dot-dashed blue line), as functions of the inter-modulus coupling $\lambda_{I}$, at single-modulus end bond $\lambda = 0.1$. Panel a): each modulus with $n=6$ sites. Panel b): each modulus with $n=7$ sites. Insets: $C_{1,n} \equiv C_{n+1,2n}$ (dotted black line), $C_{1,2} \equiv C_{2n-1,2n}$ (dashed orange line), and $(\tau^{res}_1)^{1/2}$ (dot-dashed blue line) inside noninteracting moduli ($\lambda_{I}=0$), all as functions of the single-modulus end bond $\lambda$.}
\end{figure}

{\bf Introduction:} Entanglement generation and distribution over long distances is crucial
to the realization of large-scale quantum computation and information tasks, including
teleportation, memories and repeaters, and secure key distribution \cite{QI}. Unfortunately, in systems with short-range interactions bipartite entanglement decreases rapidly with distance \cite{XY}, whereas in systems with long-range interactions entanglement monogamy prevents the creation of strongly entangled pairs \cite{monogamy}.
%%A first step in the quest for creating sufficiently large and robust
%%entanglement between distant, non directly interacting systems has been taken with the introduction of
%%localizable entanglement, a measure of the average entanglement that can be concentrated on a pair of
%%arbitrarily distant constituents by performing optimal local measurements onto the rest of the system %%\cite{Loc}. %\\
Systems with finite correlation length, such as the $1$-D Heisenberg and $XX$ models, allow for sizeable ground-state end-to-end entanglement, independent of the size of the system, provided that simple patterns of site-dependent couplings are selected. This phenomenon has been termed Long-Distance Entanglement (LDE) \cite{LDE}. It can be optimized against thermal decoherence in models with more sophisticated patterns of couplings \cite{LDEsa1}, and might be implemented using suitably engineered atom-optical systems, including optical lattices and coupled cavity arrays \cite{LDEsa1,LDEsa2}. However, its thermal instability remains a paramount obstacle against possible implementations. \\
In the present work we consider the general problem of generating and distributing entanglement between distant and non-interacting quantum systems. We discuss the properties of renormalization, amplification, and scalability of end-to-end entanglement in quantum many-body systems. From this investigation there emerges naturally the concept of {\it Modular Entanglement}. Specifically, we will show that size-independent end-to-end entanglement arises in the ground state of {\it modular} systems constituted by a set of identically interacting moduli of arbitrarily fixed size. This general type of entanglement at long distances, termed Modular Entanglement (ME), includes LDE as the particular case realized in systems 
formed of two-qubit moduli.\\
We will show that a most relevant feature of ME is its enhanced stability against thermal decoherence, even by several orders of magnitude, compared to the case of simple LDE.
%As a consequence, the relation between an alternated structure of the couplings and the onset of LDE becomes %only a feature of the two-spin modulus case.
Indeed, as we will discuss, genuine ME is generated also in systems composed by moduli that, individually, do not exhibit sizeable end-to-end entanglement. This is in sharp contrast with systems featuring LDE, for which the individual moduli always exhibit maximal end-to-end entanglement. Besides showing that LDE is a special, singular case of the much more general ME, these findings imply that end-to-end entanglement at large distances is an emerging collective property that need not be shared by the individual constituents.\\
%In particular our analysis will be focused on models with $XX$ type  interactions  \cite{Lieb,Albanese}.
%The results obtained for this kind of interactions can be straightforwardly generalized. In facts,  it has %been shown \cite{LDEsa1} that the ground state entanglement properties of XX models are fully equivalent to %those of the Heisenberg antiferromagnetic model, with the advantage of computational feasibility and of %being independent of the interaction sign  (i.e. either ferro or antiferromagnetic).
%Establishing ME sets the general requirements to be met for obtaining nonvanishing quantum correlations %between distant and non-interacting constituents, including the thermodynamic limit, and paves the way to %system optimization.
We will investigate first the onset of ME in minimal modular systems 
%%\cite{Lieb}%% 
made of two identically interacting moduli. We will then consider many-moduli systems and discuss the mechanisms associated to the onset of ME and its scaling behaviour with the number of moduli. \\
%This simple analysis allows us to point out two categories of building blocks. Whereas even site moduli %allow the onset of end-to-end quantum correlations in the composite system,  on the contrary in odd site %moduli  these correlations vanish for almost all values of the parameters.  We then analyze the %thermodynamic limit of such composite systems, showing both the generation of true LDE and the rapid %achievement of the asymptotic value.
{\bf{Two moduli:}} Here and throughout the paper we will consider $1$-D systems made of $N$ moduli each containing $n$ spins, such that the total number of spins in the system is $n_t = nN$. We first consider the most elementary modular system composed of $N=2$ moduli $B_1$ and
$B_2$ (See Fig.~\ref{Due-N-blocchi}, upper panel), endowed with site-dependent, isotropic, nearest-neighbor exchange interactions. The total Hamiltonian of the system reads:
\be  H_{T,2moduli} = H^{B_1}_{1,n} + H^{B_2}_{n+1,2n} + H^{I}_{n,n+1} \; ,
\label{2blocks}
\ee
with
\be H^\alpha_{\gamma,\delta}=\frac{1}{2}\sum_{i=\gamma}^{\delta-1}J_{i,i+1}(S^x_iS^x_{i+1}+S^y_iS^y_{i+1})
\; ,
\label{XX}
\ee
where $H^{I}_{n,n+1}$ is the interaction Hamiltonian between the two boundary qubits in $B_1$ and $B_2$.
We now investigate how and to what extent the generation of end-to-end bipartite entanglement in a
modular system depends on the entanglement properties of the individual constituent moduli.
Intuitively, we expect that the basic requirement on a single modulus is the existence of values in the set of couplings $\{J_{i,i+1}\}$ such that the end qubits in the modulus form a maximally entangled pair (a dimer).  This implies a structure of symmetric couplings inside the modulus as, in general, mirror symmetry is a necessary condition for nonvanishing end-to-end entanglement \cite{LDEsa1,Albanese,njp}. As the final yield will be qualitatively independent of the specific realization, we take as benchmark constituent modulus a uniformly interacting system with distorted end bonds: $J_{1,2}=J_{n-1,n}=\lambda J$, $J_{i,i+1}=J$ for $i\in[2,n-2]$. Fixing $J$ as the unit, one is left with two (dimensionless) free single-modulus parameters: the end bond coupling $\lambda$ and the modulus size $n$. \\
%The qualitative equivalence of all our results for different coupling realizations has been anyway checked %numerically. \\
In Fig. \ref{2blocksevodd} the overall end-to-end bipartite ME in the reduced state of the end qubits $1$ and $2n$, measured by the concurrence $C_{1,2n}$ \cite{concurrence}, is plotted as a function of the remaining free parameter, the modulus-modulus interaction coupling $\lambda_{I}$, for the most elementary modular system composed just of two identical moduli. For comparison, Fig. \ref{2blocksevodd} reports also the end-to-end entanglement and the nearest-neighbor (n.n.) end-pair concurrences inside a single modulus, as well as the multipartite entanglement measured by the residual tangle $\tau^{res}_i$ \cite{tangle}.
%It is defined by subtracting the sum of all possible two-qubit concurrences from the bipartite entanglement %of qubit $S_i$ with the rest of the system. The upper and lower panels illustrate, respectively, the %instance of moduli containing an even and an odd number of spins.
%In particular, as a measure of two site  entanglement we adopt the concurrence  which, due to the symmetries %of the  XX model, depends only on a single two-point correlation function $\langle S^{+}_iS^{-}_j\rangle$ %(where $S^{\pm}_i$ are the ladder operators on the $i$-th site of the chain) \cite{LDEsa1}.
\begin{figure}[b]
\includegraphics[width=7.5cm]{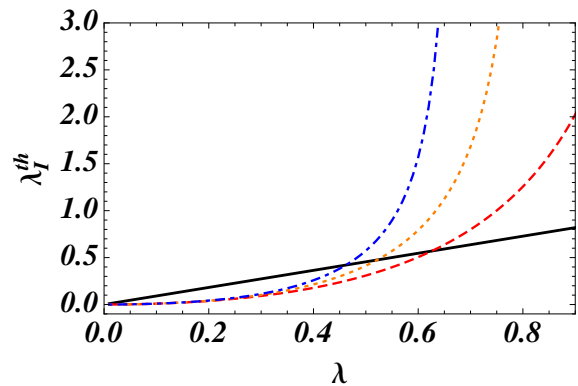}
\caption{\label{thr2} (color online) Inter-modulus threshold coupling $\lambda^{th}_I$ for the onset of ME as a function of the single-modulus end bond $\lambda$ in two-moduli systems with a number of
sites per modulus $n=2$ (solid black line), $n=4$ (dashed red line), $n=6$ (dotted orange line), and
$n=8$ (dot-dashed blue line).}
\end{figure}
In Fig. \ref{2blocksevodd} the single-modulus end bond $\lambda$ is such that, for $\lambda_{I}=0$,
both the end-to-end entanglement and the end-pair n.n. concurrences are nonvanishing inside the modulus, as shown in the insets. Panel a) shows that as $\lambda_{I}$ crosses a threshold value $\lambda_I^{th}$, the ME is nonvanishing and monotonically increasing up to a saturation value always larger than the end-to-end entanglement originally present in the noninteracting moduli ($\lambda_I = 0$). The threshold coupling $\lambda_I^{th}$ depends on the single-modulus end-to-end entanglement via the single-modulus end bond $\lambda$ and the number $n$ of sites per modulus (see Fig. \ref{thr2}).
%Moreover, the ME reaches the maximum (unit) value at small but finite values of $\lambda$.
The rise of ME is to the detriment of the end-to-end entanglement and end-pair concurrences inside each single modulus. Both rapidly vanish with increasing $\lambda_I$. On the other hand, both ME and multipartite entanglement initially increase with $\lambda_I$ until, exactly at the crossing of ME with the single-modulus end-to-end entanglement, the residual tangle peaks and then decreases for larger values of $\lambda_I$. Therefore, the onset of ME in the ground state of a two-moduli system with an even number of qubits per modulus is due to the conversion of all forms of bipartite entanglement, originally present in the non-interacting moduli, into ME and, perhaps even more strikingly, the partial conversion into ME of the ground-state multipartite entanglement.
%Essentially the
%same results hold for a two-modulus modular systems composed of even moduli of different sizes.
%The two spin modulus composite system behaves differently as, due to direct interaction, the bare modulus %concurrence is always one.
Viceversa, no ME is created in a two-moduli system with an odd number of qubits per modulus,
due to the impossibility of suppressing multipartite correlations inside each modulus, as shown in panel b) of Fig. \ref{2blocksevodd}).
%Essentially the same results hold for a two-modulus modular system composed by an even and an odd modulus.
%(see the inset in the lower panel of Fig. \ref{2blocksevodd}), the inter-modulus interaction causes the modulus %correlation to vanish without giving rise to any extremal entanglement, except for the very small region %with $\lambda_I$ close to zero (see lower panel of Fig. \ref{2blocksevodd}). In this region, we can observe %an almost negligible amount of end-to-end entanglement resulting from perturbative effects due to a %reduction in  ground state degeneracy. The double degeneracy of the non interacting odd modulus ground state %is  in fact removed by switching on the inter modulus interaction.
The presence of a region where end-to-end entanglement prevails over multipartite entanglement inside a noninteracting modulus emerges as the distinctive feature for the presence of ME in a two-moduli system.\\
%Moreover, the higher the end-to-end concurrence, the higher the contribution of multipartite correlations in %apparent contrast with the even site case. Due to symmetry, for odd site moduli entanglement localization at %the extremes minimizes the one-tangle bipartite contribution.
%We can hence conclude that if a system admits a parameter region such that the end-to-end entanglement (LDE) %is maximal or, equivalently, where the multipartite entanglement tends to be suppressed, then it can be used %as a building modulus for a composite system with nonvanishing end-to-end entanglement (ME).
\begin{figure}[t]
\includegraphics[width=8cm,height=4.5cm]{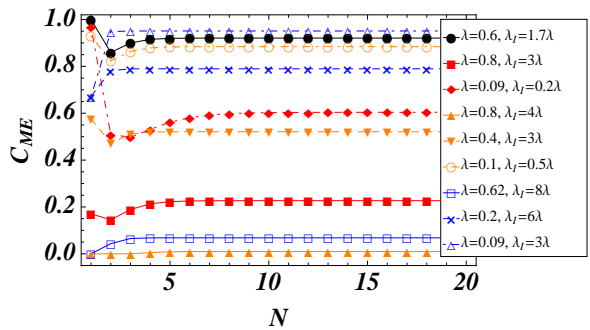}
\caption{\label{therm} (color online) End-to-end concurrence, for different values of
the intra-modulus and inter-modulus couplings $\lambda$ and $\lambda_I$, as a function of the
number of moduli $N$ in systems with a number of qubits per modulus $n=2$
(black full circles), $n=4$ (red full squares and diamonds), $n=6$  (orange full triangles and
inverted triangles, and empty circles), and $n=8$ (blue empty triangles and squares, and
dashed line). We consider systems up to $n_t = 160$ qubits, corresponding
to $N=20$ moduli and $n=8$ qubits per modulus.}
\end{figure}
\begin{figure}[b]
\includegraphics[width=8cm,height=4.5cm]{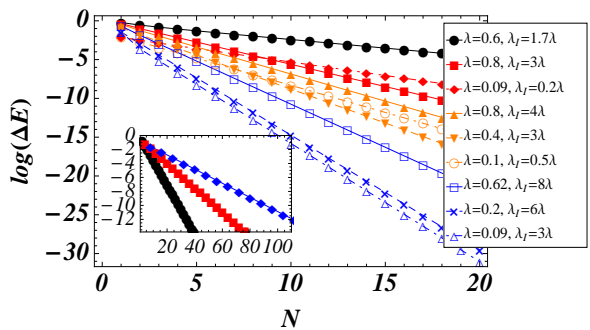}
\caption{\label{therm2} (color online) Energy gap, for different values of
the intra-modulus and inter-modulus couplings $\lambda$ and $\lambda_I$,
as a function of the number of moduli $N$ in systems with a number of
spins per modulus $n=2$ (black full circles), $n=4$ (red full squares and diamonds),
$n=6$ (orange full triangles and inverted triangles, and empty circles), and
$n=8$ (blue empty triangles and squares, and dashed line).
Inset: energy gap as a function of the total number of sites $Nn$ in systems
with a number of spins per modulus $n=2$ (black full circles), $n=4$ (red full squares),
and $n=8$ (blue full diamonds). As in Fig. \ref{therm}, we consider 
systems up to $n_t = 160$ spins.}
\end{figure}
\begin{figure}[t]
\includegraphics[width=8cm,height=4.5cm]{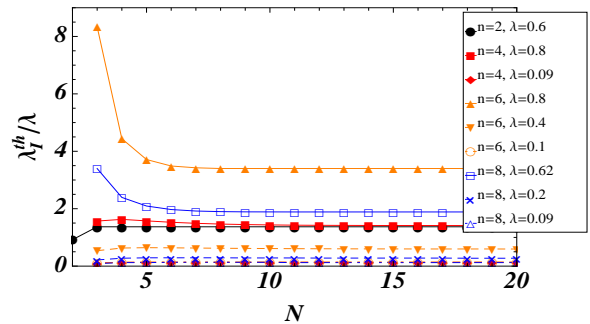}\caption{\label{thr1} (color online) Modulus-modulus threshold coupling $\lambda^{th}_I$ for the onset of ME as a function of
the number of moduli $N$, in a system with a number of spins per modulus $n=2$
(black full circles), $n=4$ (red full squares and diamonds), $n=6$ (orange full triangles and
inverted triangles), and $n=8$ (blue empty triangles and squares, and dashed line). As in Fig. \ref{therm}, we consider systems up to $n_t = 160$ spins.}
\end{figure}
{\bf Many moduli:} We now investigate systems made of an arbitrary number of interacting moduli.
We will show that such systems exhibit genuine ME, i.e. size-independent, nonvanishing end-to-end entanglement. We consider the class of Hamiltonians:
\be
\sum_{k=1}^{N}H_{(k-1)n+1,kn}+\sum_{k=1}^{N-1}H^I_{kn,kn+1},
\label{nblocks}
\ee
where $H_{(k-1)n,kn}$, the Hamiltonian of the $k$th  constituting modulus replica, and $H^I_{nk,nk+1}$, the interaction Hamiltonian between the $k$th and the $(k+1)$th replicas, are both of the form Eq. (\ref{XX}), $N$ is the number of moduli and $\lambda_I$ is the inter-modulus coupling. The main result of the following
analysis will be that whenever a quantum system exhibits a structure as that of Eq. (\ref{nblocks}), then the two end-site elementary subsystems tend to form a long-distance maximally entangled pair in the ground state. The ME expressed by the end-to-end concurrence is reported in Fig. \ref{therm} as a function of the number of moduli $N$, for different values of the inter-modulus coupling $\lambda_{I}$, the number of qubits per modulus $n$, and the intra-modulus coupling $\lambda$. The data clearly show the onset of ME as well as its fast convergence to an asymptotic value. It is important to observe that, depending on the value of $\lambda_I$, the generated ME can be either larger or smaller compared to the pre-existing end-to-end entanglement inside a non-interacting modulus. Remarkably, even starting from individual moduli with vanishing end-to-end entanglement, there exist intervals of values of $\lambda_{I}$ such that a nonvanishing end-to-end ME is created by means of a collective interaction among the moduli (see e.g. the curve for a system made of six-qubit moduli with $\lambda=0.8$, $\lambda_I = 4\lambda$ in Fig. \ref{therm}). 
%%This implies that overall end-to-end entanglement can be generated even in a modular
%%system whose constituent moduli individually do not reach the threshold for the generation of {\bf{\em ETE %%entanglement.}}
Moreover, the rate of convergence to the asymptotic value of ME depends essentially on the size of the
constituent moduli. The smaller the constituent moduli, the larger the number of moduli that are needed to reach it. Furthermore (see also the discussion below), ME is generated also if $\lambda_I/\lambda<1$, at variance with the case of LDE. Indeed, in the latter case the end bonds must always be weakest \cite{LDE,LDEsa1}.
Finally, ME arises due to the interaction between moduli rather than between individual qubits. This  becomes apparent from Fig. \ref{therm2}, where the energy gap is plotted as a function of the number of moduli. In fact, the characteristic exponentially decreasing gap behavior of ME is clearly dependent on the number of moduli and not on the number of sites. In the inset of Fig. \ref{therm2} the gap is plotted as a function of the number of sites for modular systems with different numbers of sites per modulus, in the same correlation regime. Given the same total number of qubits, the energy gap in a modular system with $2n$ sites per modulus is approximately given by the square root of the gap in a modular system with $n$ sites per modulus, and so on. Therefore the ground-state ME in a modular system is much more robust against temperature than the corresponding LDE counterpart, with an improvement in the gap that can reach some orders of magnitude.
%As a side remark we note that according to whether one looks at the energy gap as a function of the number %of moduli or of sites, the  two site modulus composite system  from the most convenient  becomes the worst %one.
%The composite system energy gap $\Delta E$ can thus be estimated as
%\be \Delta E\propto\Delta\epsilon f(\lambda_{I})\exp(-N),\label{delta}\ee
%where $\Delta\epsilon$ is the non interacting modulus energy gap, $f(\lambda_I)$ is some slowly drecreasing function of the inter-modulus coupling and $N$ is the number of moduli. \\
Figs. \ref{thr2} and \ref{therm} clarify the relation between the onset of ME and the inter-modulus coupling $\lambda_I$. Whereas a nonvanishing LDE inside a single modulus generally requires a constraint on
the relative ratios of the couplings {\it within} the modulus, the collective interaction leading to the
onset of ME does not. Comparing how ME arises in systems made of moduli of equal size $n$ but different values of $\lambda$, there emerges the existence of a threshold value of $\lambda_I^{tr}$ which can be either larger or smaller than the intra-modulus coupling $\lambda$. In Fig. \ref{thr1} we plot the ratio $\lambda^{th}_I/\lambda$ as a function of the number of moduli $N$, for moduli with different $n$ and $\lambda$. By examining the curves at fixed $n$, we observe three different behaviors of $\lambda^{tr}_I$, according to the different intra-modulus correlation regimes. Namely, if the end-to-end entanglement inside each modulus is either vanishing or nearly vanishing, then the thermodynamic limit is monotonically approached from above and $\lambda_I/\lambda\gg1$; in the opposite regime, i.e. for maximal end-to-end entanglement inside the individual moduli, the limit is monotonically approached from below with $\lambda_I/\lambda\ll1$. In the intermediate regime, we observe an almost constant averaging behavior of $\lambda^{th}_I$ in the neighborhood of $\lambda$, with amplitude oscillations that slow down in the proximity of the thermodynamic value. Systems composed by minimal two-qubit moduli, i.e. those exhibiting LDE, have thus a rather pathological behavior due to the bare intra-modulus entanglement being always maximal regardless of the value of $\lambda$. Namely, the ratio $\lambda_I/\lambda$ must always be larger than unity and acquire the same for every $\lambda$.\\
%%Summing up, 
%%we have showed that there exists an ample class of quantum spin models whose ground state exhibits sizeable %%(and in many cases very large) ME, i.e. bipartite end-to-end entanglement in pairs of arbitrarily distant %%and non-interacting qubits. 
%%Summing up, the concept of ME 
%%generalizes in a nontrivial way the previously known results on Long Distance Entanglement (LDE) and 
%%allows an articulated understanding of entanglement generation and distribution in quantum spin systems.
Summing up, the existence of large end-to-end ME relies on a mechanism of balancing/compensation between the bipartite and the multipartite entanglement present in a system, as well as on the form of the energy spectrum and the degree of modularity that can be engineered in the system. One of the most important practical advantages of ME over LDE is that the energy gap scales exponentially with the total number of moduli, rather than with the total number of qubits. This difference implies that, given a fixed total number of qubits in the system, the stability of ME against thermal noise can be enhanced by orders of magnitude, compared to LDE, just by a proper tailoring of the number of moduli and of the single-modulus parameters, as shown in Fig. \ref{therm2} above. Moreover, being a static, ground-state property, there is no question on the time-scale and speed at which ME can be created. These two aspects (enhanced thermal stability and exact ground state property) of ME are crucial advantages over schemes that suggest to create entanglement between distant qubits dynamically by, e.g., a sudden global quench much below a certain threshold temperature \cite{Wichterich}. Indeed, dynamical schemes are intrinsically fragile to noise, imperfections, and thermal fluctuations. Moreover, they decay quickly with the size of the system. On the contrary, ME is a size-independent, exact ground-state property resilient to thermal decoherence and robust against imperfections. Its experimental realization and control would thus provide a relevant step towards the implementation of faithful large-scale quantum teleportation, memories, and repeaters, as well as the preparation of large multi-partite states for measurement-based quantum computation.\\
{\bf Acknowledgements:} We acknowledge financial support from the European Union under the FP7 STREP Project HIP (Hybrid Information Processing), Grant Agreement No. 221889.

\end{document}